\documentclass[graybox, envcountchap]{svmult}

\usepackage{mathptmx}        
\usepackage{amsmath}
\usepackage{amssymb}
\usepackage{color}
\usepackage{helvet}          
\usepackage{courier}         
\usepackage{dirtree}

\usepackage{makeidx}        
\usepackage{graphicx}        
\usepackage{subfig}

\usepackage{multicol}        
\usepackage[bottom]{footmisc}

\usepackage[misc]{ifsym}

\newcommand{\be}{\begin{eqnarray}}
\newcommand{\ee}{\end{eqnarray}}

\begin{document}

\title*{Fundamental Concepts}
\author{Cosimo Bambi and Sourabh Nampalliwar}
\institute{Cosimo Bambi (\Letter)
\at Department of Physics, Fudan University, 2005 Songhu Road, Shanghai 200438, China, \email{bambi@fudan.edu.cn} \and Sourabh Nampalliwar \at Theoretical Astrophysics, Eberhard-Karls Universit\"at T\"ubingen, Auf der Morgenstelle 10, 72076 T\"ubingen, Germany, \email{sourabh.nampalliwar@uni-tuebingen.de}}
%
%
\maketitle

\abstract{This chapter briefly discusses the fundamental properties of black holes in general relativity, the discovery of astrophysical black holes and their main astronomical observations, how X-ray and $\gamma$-ray facilities can study these objects, and ends with a list of open problems and future developments in the field.}


\section{Introduction}\label{s1-intro}
Beginning with the special theory of relativity in 1905, Albert Einstein soon realized that Newton's theory of gravity had to be superseded, to harmonize the equivalence principle and the special theory of relativity. After numerous insights, false alarms, and dead ends, the theory of general relativity was born in 1915~\cite{r-intro-ein}.  It took some years for it to take over Newton's theory as the leading framework for the description of gravitational effects in our Universe, and over the past century, it has become one of the bedrocks of modern physics. 

Just a year after its proposition, Karl Schwarzschild was able to find an exact solution in general relativity, much to the surprise of Einstein himself, who only had approximate solutions by that time. The Schwarzschild solution~\cite{r-intro-sch} turned out to be much more astrophysically relevant than anyone could have imagined, and describes the simplest class of \emph{black holes}\footnote{The origin of the term black hole is quite intriguing. While it is not clear who used the term first, it appeared for the first time in a publication in the January 18, 1964 issue of Science News Letter. It was on a report on a meeting of the American Association for the Advancement of Science by journalist Ann Ewing. The term became quickly very popular after it was used by John Wheeler at a lecture in New York in 1967.} in Einstein's theory. 
 

Roughly speaking, a black hole is a region in which gravity is so strong that nothing, not even light, can escape. A boundary, known as the \emph{event horizon}, separates the interior of the black hole from the exterior region and acts as a one-way membrane: particle and radiation can enter the black hole but cannot exit from it. Remarkably, a primitive concept of black hole was already discussed at the end of the 18th century in the context of Newtonian mechanics by John Michell and Pierre-Simon Laplace. The starting point was the corpuscular theory of light developed in the 17th century. Here light is made of small particles traveling with a finite velocity, say $c$. Michell and Laplace noted that the escape velocity from the surface of a body of mass $M$ and radius $R$ exceeds $c$ if $R < R_{\rm crit}$, where
\be
R_{\rm crit} = \frac{2 G_{\rm N} M}{c^2}
\ee
and $G_{\rm N}$ is Newton's gravitational constant. If such a compact object were to exist, it should not be able to emit radiation from its surface and should thus look black. This was the conclusion of Michell and Laplace and these objects were called dark stars.


The Schwarzschild type black holes are described by just one parameter, the \emph{mass}, and it is the characteristic quantity setting the size of the system. The {\it gravitational radius} of an object of mass $M$ is defined as
\be
r_{\rm g} &=& \frac{G_{\rm N} M}{c^2} 
= 14.77 \left( \frac{M}{10 \; M_\odot} \right) \text{ km } \, .
\ee
The associated characteristic time scale is
\be
\tau &=& \frac{r_{\rm g}}{c}
= 49.23 \left( \frac{M}{10 \; M_\odot} \right) \text{ $\mu$s } \, .
\ee
For a 10~$M_\odot$ black hole, $r_{\rm g} \sim 15$~km and $\tau \sim 50$~$\mu$s. We can thus expect that physical phenomena occurring around a similar object can have a variability timescale of the order of 0.1-1~ms. For a black hole with $M \sim 10^6$~$M_\odot$, we find $r_{\rm g} \sim 10^6$~km and $\tau \sim 5$~s, so physical processes occurring near its gravitational radius can have a variability timescale of the order of 10-100~s. For the most supermassive black holes with $M \sim 10^9$~$M_\odot$, we have $r_{\rm g} \sim 10^9$~km and $\tau \sim 1$~hr. 

The astrophysical implications of such black hole solutions were not taken very seriously for a long time. For example, influential scientists like Arthur Eddington argued that ``some unknown mechanism'' had to prevent the complete collapse of a massive body and the formation of a black hole in the Universe. The situation changed only in the 1960s with the advent of X-ray observations. Yakov Zel’dovich and, independently, Edwin Salpeter were the first, in 1964, to propose that quasars were powered by central supermassive black holes~\cite{r-intro-zeld,r-intro-salp}. In the early 1970s, Thomas Bolton and, independently, Louise Webster and Paul Murdin identified the X-ray source Cygnus X-1 as the first stellar-mass black hole candidate~\cite{r-intro-cyg2,r-intro-cyg3}. The uncertainty of those times can be imagined by the scientific wager between Kip Thorne and Stephen Hawking, the latter claiming that Cygnus X-1 was in fact not a black hole. Hawking conceded the bet in 1990. In the past few decades, a large number of astronomical observations have pointed out the existence of stellar-mass black holes in some X-ray binaries~\cite{r-intro-re-mc} and of supermassive black holes at the center of many galaxies~\cite{r-intro-k-r}. Thanks to X-ray and $\gamma$-ray missions like \textsl{XMM-Newton}, \textsl{Chandra}, \textsl{NuSTAR}, \textsl{Swift}, and \textsl{Fermi}, in the past 20~years there have been substantial progresses in the study of these objects. In September~2015, the LIGO experiment detected, for the first time, the gravitational waves emitted from the coalescence of two black holes~\cite{r-intro-gw150914}.



\section{Black holes in general relativity}\label{s1-gr}

In 4-dimensional general relativity, black holes are relatively simple objects, in the sense that they are completely characterized by a small number of parameters: the mass $M$, the spin angular momentum $J$, and the electric charge $Q$. This is the result of the {\it no-hair theorem}, which holds under specific assumptions~\cite{r-intro-h1,r-intro-h2,r-intro-h3,r-intro-h4}. The name ``no-hair'' refers to the fact black holes have only a small number of features (hairs). Violations of the no-hair theorem are possible in the presence of exotic fields, extra dimensions, or extensions of general relativity.

A {\it Schwarzschild black hole} is a spherically symmetric, non-rotating, and electrically uncharged black hole and is completely characterized by its mass. In the presence of a non-vanishing electric charge, we have a {\it Reissner-Nordstr\"om black hole}, which is completely specified by two parameters and describes a spherically symmetric and non-rotating black hole of mass $M$ and electric charge $Q$. A {\it Kerr black hole} is an uncharged black hole of mass $M$ and spin angular momentum $J$. The general case is represented by a {\it Kerr-Newman black hole}, which has a mass $M$, a spin angular momentum $J$, and an electric charge $Q$.

Astrophysically, black holes are expected to belong to the Kerr family. After the collapse of a massive body and the creation of an event horizon, the gravitational field of the remnant quickly reduces to that of a Kerr black hole by emitting gravitational waves~\cite{r-intro-k1,r-intro-k2}. For astrophysical macroscopic objects, the electric charge is extremely small and can be ignored~\cite{r-intro-k3,r-intro-book}. The presence of an accretion disk around the black hole, as well as of stars orbiting the black hole, do not appreciably change the strong gravity region around the compact object~\cite{r-intro-k4,r-intro-k5,r-intro-k6}. Astrophysical black holes should thus be completely specified by their mass and spin angular momentum. It is often convenient to use the dimensionless spin parameter $a_*$ instead of $J$. For a black hole of mass $M$ and spin $J$, $a_*$ is defined as
\be
a_* = \frac{c J}{G_{\rm N} M^2} \, .
\ee

In general relativity, the choice of the coordinate system is arbitrary, and therefore the numerical values of the coordinates have no physical meaning. Nevertheless, they can often provide the correct length or time scale of the system. In Boyer-Lindquist coordinates, the typical coordinate system for Kerr black holes, the radius of the event horizon is
\be\label{eq-horizon}
r_{\rm H} = r_{\rm g} \left( 1 + \sqrt{1 - a_*^2} \right) \, ,
\ee
and depends on $M$ (via $r_{\rm g} $ and $a_*$) and $J$ (via $a_*$). The radius of the event horizon thus ranges from $2 \, r_{\rm g}$ for a non-rotating black hole to $r_{\rm g}$ for a maximally rotating ($a_* = \pm 1$) black hole. Note that Eq.~(\ref{eq-horizon}) requires that $| a_* | \le 1$. Indeed for $| a_* | > 1$ there is no black hole and the Kerr solution describes the gravitational field of a \emph{naked singularity}. In the context of astrophysical observations, the possibility of the existence of naked singularity is usually ignored, and this is also motivated by the considerations that $i)$ there is no known mechanism capable of creating a naked singularity, and $ii)$ even if created, the spacetime is likely unstable (for more details, see for instance Ref.~\cite{r-intro-book}).

\begin{figure}[t]
\begin{center}
\includegraphics[width=10.0cm]{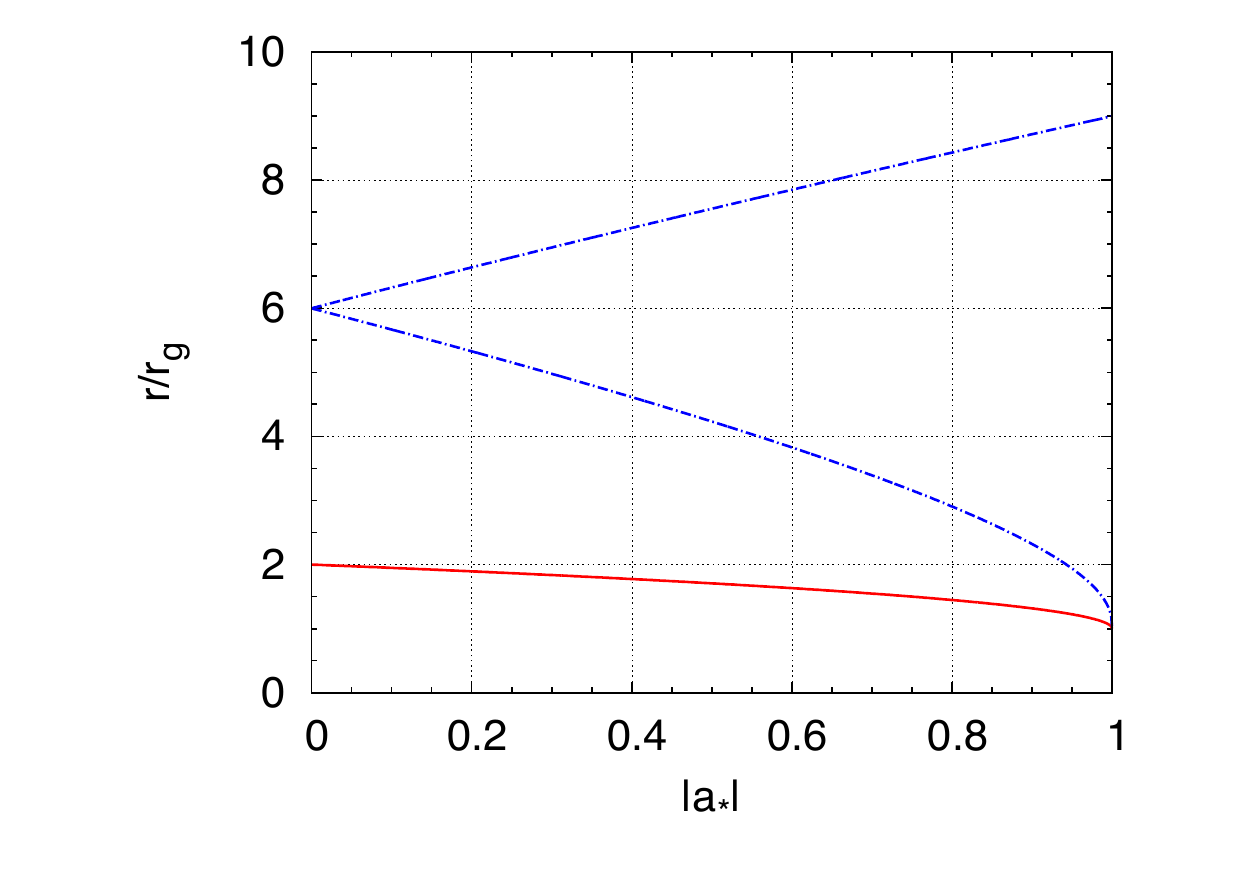}
\end{center}
\vspace{-0.6cm}
\caption{Radius of the event horizon (red solid line) and of the ISCO (blue dash-dotted line) of a Kerr black hole in Boyer-Lindquist coordinates as a function of the spin parameter $a_*$. For the ISCO radius, the upper curve refers to counterrotating orbits and the lower curve to corotating orbits. \label{f-isco}}
\end{figure}

The properties of equatorial circular orbits around a black hole are important for astrophysical observations because they describe the orbits of the particles in a putative accretion disk around the compact object. In Newtonian mechanics, equatorial circular orbits (i.e. orbits in the plane perpendicular to the spin of the object) around a point-like object are always stable. However, this is not true for equatorial circular orbits around a Kerr black hole. Here we have the existence of an {\it innermost stable circular orbit}, often abbreviated to ISCO. In Boyer-Lindquist coordinates, the ISCO radius is~\cite{r-intro-bpt72} 
\be\label{eq-intro-isco}
r_{\rm ISCO} &=& r_{\rm g} \left[ 3 + Z_2 \mp \sqrt{\left(3 - Z_1\right)
\left(3 + Z_1 + 2 Z_2\right)} \right] \, , \nonumber\\
Z_1 &=& 1 + \left(1 - a^2_*\right)^{1/3}
\left[\left(1 + a_*\right)^{1/3} + \left(1 - a_*\right)^{1/3}\right]\, , \nonumber\\
Z_2 &=& \sqrt{3 a^2_* + Z^2_1} \, .
\ee
The ISCO radius turns out to be $6 \, r_{\rm g}$ for a Schwarzschild black hole and move to $r_{\rm g}$ ($9 \, r_{\rm g}$) for a maximally rotating black hole and a corotating (counterrotating) orbit, namely an orbit with angular momentum parallel (antiparallel) to the black hole spin. Fig.~\ref{f-isco} shows the radial values of the event horizon $r_{\rm H}$ and of the ISCO radius $r_{\rm ISCO}$ in Boyer-Lindquist coordinates as a function of the black hole spin parameter $a_*$.

\section{Black holes in astrophysics}\label{s1-astro}

While we cannot observe any kind of radiation (neither electromagnetic, nor gravitational) from the region inside the event horizon, astrophysical black holes can be studied by detecting the electromagnetic and gravitational radiation produced in the vicinity of the event horizon. Gravitational radiation is generated by the interaction of matter/energy and the spacetime, and its frequency depends on the size of the system. In particular, the wavelength roughly scales as the linear size of the system emitting gravitational radiation. Gravitational radiation from black holes is expected to range from a few nHz, in the case of the merger of galaxies with supermassive black holes at their respective centers, to a few kHz, in the case of the merger and ringdown of stellar-mass black holes. Radiation of different wavelengths require different observational facilities to be detected.

\begin{figure}[h]
\vspace{0.5cm}
\begin{center}
\includegraphics[width=10.0cm]{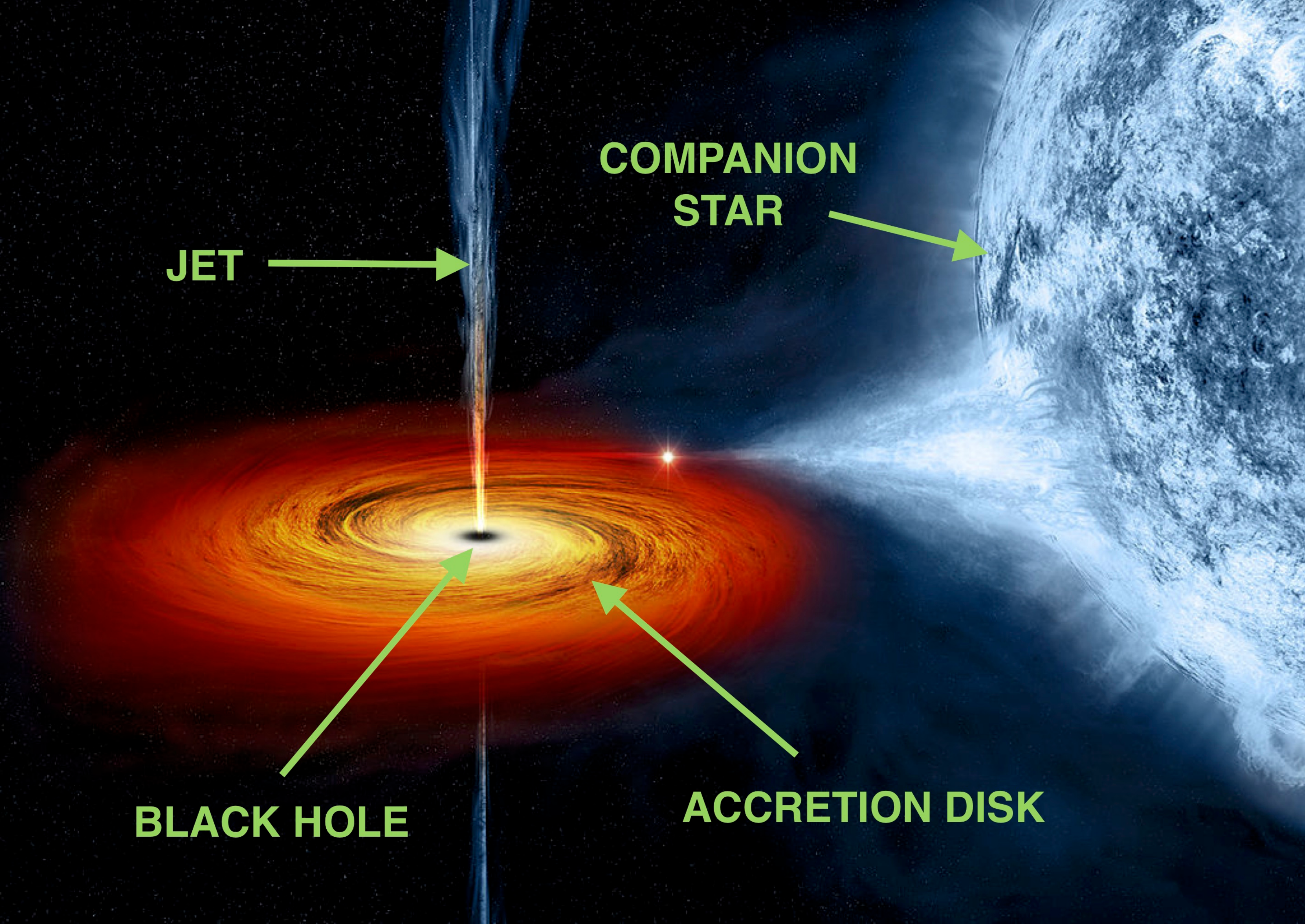}
\end{center}
\vspace{0.3cm}
\caption{An artist's illustration of Cygnus~X-1. The stellar-mass black hole pulls material from a massive, blue companion star toward it. This material forms an accretion disk around the black hole. We also see a jet originating from the region close to the black hole. Credit: NASA. \label{f-art}}
\end{figure}

Electromagnetic radiation can be emitted by the gas in the accretion disk, jet, and outflows, as well as by possible bodies (like stars) orbiting the black hole (see Fig.~\ref{f-art}). The electromagnetic spectra of astrophysical black holes range from the radio to the $\gamma$-ray band (see Tab.~\ref{of-tab-band} for the list of the bands of the electromagnetic spectrum). The photon energy is determined by the emission mechanism and the black hole environment. Photons with different wavelengths carry different information about the black hole and its environment, and require different observational facilities to be detected. Tab.~\ref{of-tab-components} lists the possible components of the electromagnetic spectrum of a black hole system (more details on each component will be provided in the next chapter).

\begin{table}[h]
\centering
{\renewcommand{\arraystretch}{1.5}
\begin{tabular}{|lccc|}
\hline
$\hspace{0.1cm}$ Band & Wavelength & Frequency & Energy \\
\hline
$\hspace{0.1cm}$ Radio & $> 0.1$~m & $< 3$~GHz & $< 12.4$~$\mu$eV \\
$\hspace{0.1cm}$ Microwave & 1~mm-0.1~m & 3-300~GHz & $\hspace{0.2cm}$ 12.4~$\mu$eV-1.24~meV $\hspace{0.2cm}$ \\
$\hspace{0.1cm}$ Infrared (IR) & $\hspace{0.2cm}$ 700~nm-1~mm $\hspace{0.2cm}$ & 300~GHz-430~THz & 1.24~meV-1.7~eV \\
$\hspace{0.1cm}$ Visible & 400-700~nm & 430-790~THz & 1.7-3.3~eV \\
$\hspace{0.1cm}$ Ultraviolet (UV) $\hspace{0.2cm}$ & 10-400~nm & $\hspace{0.2cm}$ $7.9 \cdot 10^{14}$-$3 \cdot 10^{16}$~Hz $\hspace{0.2cm}$ & 3.3-124~eV \\
$\hspace{0.1cm}$ X-Ray & 0.01-10~nm & $3 \cdot 10^{16}$-$3 \cdot 10^{19}$~Hz & 124~eV-124~keV \\
$\hspace{0.1cm}$ $\gamma$-Ray & $< 0.01$~nm & $> 3 \cdot 10^{19}$~Hz & $> 124$~keV \\
\hline
\end{tabular}} 
\vspace{0.4cm}
\caption{Bands of the electromagnetic spectrum. Note that different authors may use slightly different definitions. \label{of-tab-band}}
\end{table}

\begin{table}[h]
\centering
{\renewcommand{\arraystretch}{1.5}
\begin{tabular}{|lccc|}
\hline
$\hspace{0.1cm}$ Source $\hspace{0.1cm}$ & $\hspace{0.1cm}$ Emission $\hspace{0.1cm}$ & $\hspace{0.1cm}$ Stellar-mass black holes $\hspace{0.1cm}$ & $\hspace{0.1cm}$ Supermassive black holes $\hspace{0.1cm}$ \\
\hline
$\hspace{0.1cm}$ Accretion disk & Thermal & UV to soft X-ray & Visible to UV \\
$\hspace{0.1cm}$ Accretion disk & $\hspace{0.1cm}$ Reflection spectrum $\hspace{0.1cm}$ & X-ray & X-ray \\
$\hspace{0.1cm}$ Corona & Inverse Compton & X-ray and $\gamma$-ray & X-ray and $\gamma$-ray \\
$\hspace{0.1cm}$ Jet & Synchrotron & Radio to soft X-ray & Radio to soft X-ray \\
$\hspace{0.1cm}$ Jet & Inverse Compton & X-ray and $\gamma$-ray & X-ray and $\gamma$-ray \\
$\hspace{0.1cm}$ Cold material & Emission lines & --- & IR to X-ray \\
$\hspace{0.1cm}$ Companion star & Thermal & Visible and UV & --- \\
\hline
\end{tabular}} 
\vspace{0.4cm}
\caption{Summary of the possible sources of electromagnetic radiation in black hole systems and typical energy bands for stellar-mass and supermassive black holes. For soft X-ray we mean the X-ray band below a few keV. Cold material orbiting the compact object and not belonging to the accretion disk is common in supermassive black holes: the emission lines can be narrow (broad) if the material is far (near) the compact object and moving with low (high) speed. \label{of-tab-components}}
\end{table}

\begin{figure}[t]
\begin{center}
\includegraphics[width=10.0cm]{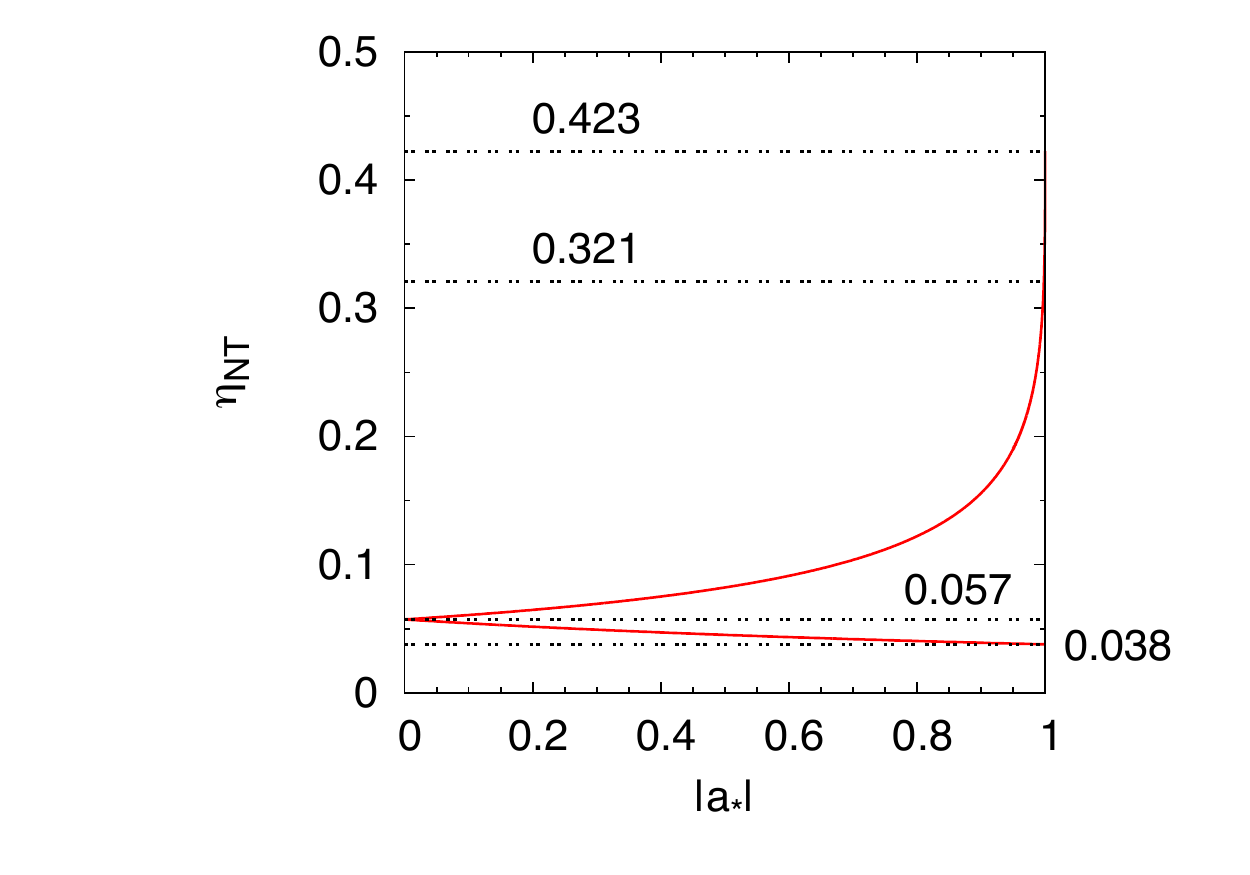}
\end{center}
\vspace{-0.6cm}
\caption{Efficiency of a Novikov-Thorne disk $\eta_{\rm NT}$ as a function of the spin parameter $a_*$ for Kerr black holes. The upper (lower) curve is for corotating (counterrotating) orbits. The dotted horizontal lines mark the radiative efficiencies for $a_* = 1$ ($\eta_{\rm NT} \approx 0.423$), $a_* = 0.998$ ($\eta_{\rm NT} \approx 0.321$), $a_* = 0$ ($\eta_{\rm NT} \approx 0.057$), and $a_* = -1$ ($\eta_{\rm NT} \approx 0.038$). \label{f-eta-NT}}
\end{figure}

Among the various astrophysical processes, accretion onto a black hole can be an extremely efficient mechanism to convert mass into energy. If $\dot{M}$ is the mass accretion rate, the total power of the accretion process can be written as
\be
P = \eta \dot{M} c^2 \, ,
\ee
where $\eta$ is the total efficiency. In general, the energy released in the accretion process will be converted into radiation and kinetic energy of jets/outflows, so we can write
\be
\eta = \eta_{\rm r} + \eta_{\rm k} \, ,
\ee
where $\eta_{\rm r}$ is the radiative efficiency and can be measured from the bolometric luminosity $L_{\rm bol}$ from the equation $L_{\rm bol} = \eta_{\rm r} \dot{M} c^2$ if the mass accretion rate is known, and $\eta_{\rm k}$ is the fraction of gravitational energy converted into kinetic energy of jets/outflows. The actual efficiency depends on the morphology of the accretion flow. In the case of a Novikov-Thorne disk (see next chapter), the accretion disk is on the black hole equatorial plane, perpendicular to the spin of the compact object. The particles of the gas follow equatorial circular orbits, they lose energy and angular momentum, and they move to smaller and smaller radii. When the particles reach the ISCO radius, they quickly plunge onto the black hole, without significant emission of additional radiation. The efficiency of the process is thus given by
\be
\eta_{\rm NT} = 1 - E_{\rm ISCO} \, ,
\ee
where $E_{\rm ISCO}$ is the specific energy of the gas at the ISCO radius, namely the energy per unit mass of the gas. For a Kerr black hole, the specific energy of a particle orbiting an equatorial circular orbit at the Boyer-Lindquist radial coordinate $r$ is~\cite{r-intro-book} 
\be
E = \frac{r^{3/2} - 2 r_{\rm g} r^{1/2} \pm a_* r_{\rm g}^{3/2}}{r^{3/4} 
\sqrt{r^{3/2} - 3 r_{\rm g} r^{1/2} \pm 2 a_* r_{\rm g}^{3/2}}} \, .
\ee
If we plug the radial coordinate of the ISCO radius in Eq.~(\ref{eq-intro-isco}), we find that the efficiency of the process is around 5.7\% for a Schwarzschild black hole and monotonically increases (decreases) as the spin parameter increases (decreases) up to about 42.3\% (3.8\%) for $a_* = 1$ ($a_* = -1$):
\be
\eta_{\rm NT} (a_* = 0) &=& 1 -  \frac{2 \sqrt{2}}{3} \approx 0.057 \, , \nonumber\\
\eta_{\rm NT} (a_* = 1) &=& 1 -  \frac{1}{\sqrt{3}} \approx 0.423 \;\;\; 
({\rm corotating \;\; disk}) \, , \nonumber\\
\eta_{\rm NT} (a_* = - 1) &=& 1 -  \frac{5}{\sqrt{27}} \approx 0.038 \;\;\; 
({\rm counterrotating \;\; disk}) \, .
\ee
Fig.~\ref{f-eta-NT} shows $\eta_{\rm NT}$ as a function of the spin parameter $a_*$ for corotating (upper curve) and counterrotating (lower curve) disks. The efficiency of a Novikov-Thorne disk can be compared to other astrophysical processes. For instance, if we consider nuclear reactions inside the Sun, the main process is the fusion of protons to form helium-4 nuclei. The total mass of the final state is lower than the total mass of the initial state, and this difference is released into energy (electromagnetic radiation and kinetic energy of the particles in the final state). The efficiency of the process is only around 0.7\%, namely about 0.7\% of the initial mass is converted into energy.

If the mass accretion rate is low and the accreting gas has a low angular momentum, the efficiency of the accretion process can be much smaller than 1, $\eta \ll 1$, because the particles of the gas simply fall onto the gravitational well of the black hole without releasing much electromagnetic radiation. Very low efficiencies are also possible in the case of very high mass accretion rate, and in this case it is because the particle density of the accretion flow is too high and the medium becomes optically thick to the radiation emitted by the gas, so everything is advected onto the black hole and lost after crossing the event horizon. An important concept in this regard is the {\it Eddington luminosity}. The concept is actually more general, and the Eddington luminosity refers to the maximum luminosity for an object, not necessarily a black hole. The Eddington luminosity $L_{\rm Edd}$ is reached when the pressure of the radiation luminosity on the emitting material balances the gravitational force towards the object. If a normal star has a luminosity $L > L_{\rm Edd}$, the pressure of the radiation luminosity drives an outflow. If the luminosity of the accretion flow of a black hole exceeds $L_{\rm Edd}$, the pressure of the radiation luminosity stops the accretion process, reducing the luminosity. Assuming that the emitting medium is a ionized gas of protons and electrons, the Eddington luminosity of an object of mass $M$ is
\be
L_{\rm Edd} = \frac{4 \pi G_{\rm N} M m_p c}{\sigma_{\rm Th}}
= 1.26 \cdot 10^{38} \left(\frac{M}{M_\odot}\right) \text{ erg/s} \, ,
\ee
where $m_p$ is the proton mass and $\sigma_{\rm Th}$ is the electron Thomson cross section.  For an accreting black hole, we can define the Eddington mass accretion rate $\dot{M}_{\rm Edd}$ from
\be\label{eq-intro-mdotedd}
L_{\rm Edd} = \eta_{\rm r} \dot{M}_{\rm Edd} c^2 \, , 
\ee
where $\eta_{\rm r}$ is still the radiative efficiency.


\section{X-ray and $\gamma$-ray observatories}\label{s1-obs}

\begin{figure}[t]
\vspace{0.8cm}
\begin{center}
\includegraphics[width=12cm]{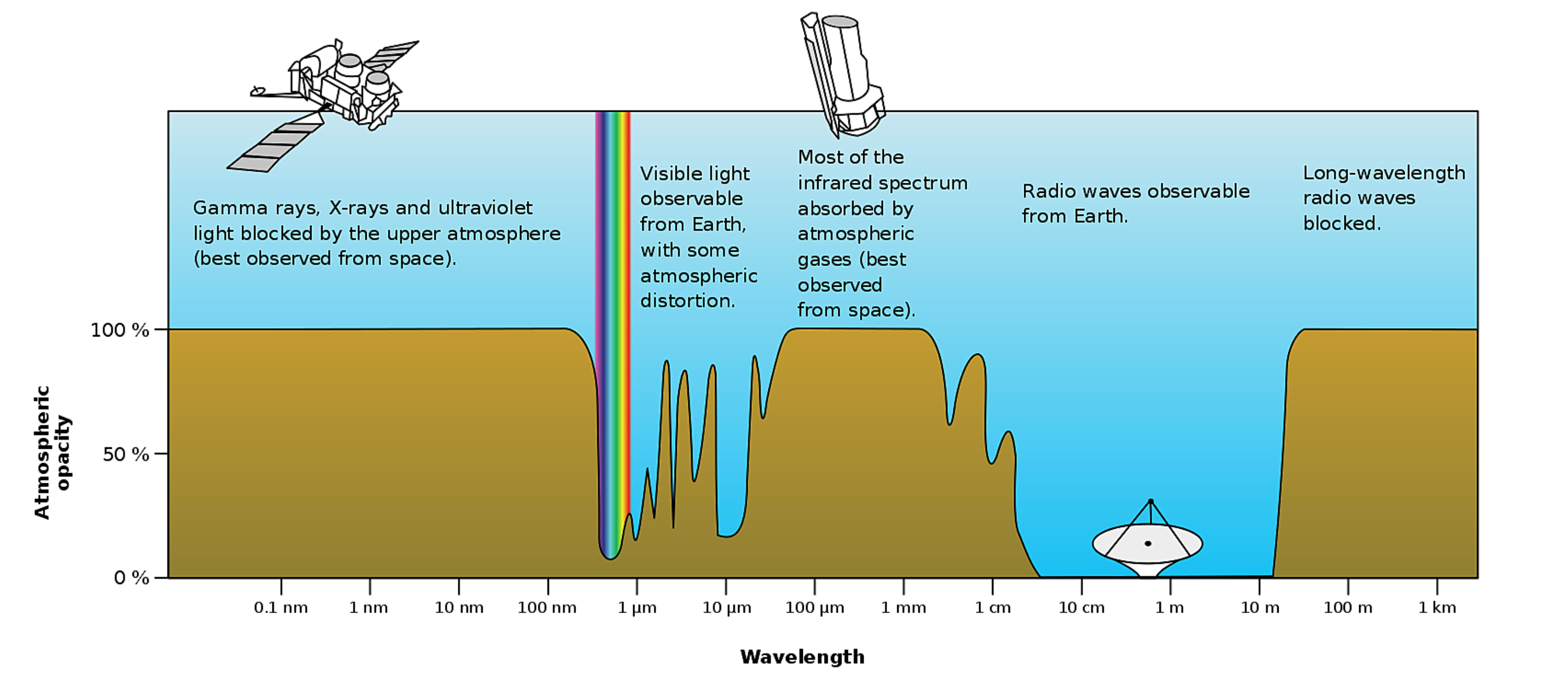}
\end{center}
\vspace{-0.1cm}
\caption{Atmospheric opacity as a function of photon wavelength. Since the atmosphere is opaque at most wavelengths, only optical and radio telescopes can be at ground level on Earth. $\gamma$-ray, X-ray, UV, and IR observational facilities are required to be on board of rockets or satellites. Credit: NASA. \label{f-atmosphere}}
\end{figure}

Our focus in this book is X-rays and $\gamma$-rays. There are a number of astrophysical sources emitting X-ray (0.1-100~keV) and $\gamma$-ray ($>100$~keV) radiation, such as galaxy clusters, compact objects, supernova remnants, and stars. X-ray radiation can be emitted by hot gas (10$^6$-10$^9$~K) or generated by bremsstrahlung, synchrotron processes, inverse Compton scattering, fluorescent emission, and nuclear decay. $\gamma$-ray radiation can be generated by the same processes at higher energies, as well as by electron-positron annihilation. As a back-of-the-envelope estimate, consider an electromagnetic particle falling onto a black hole, beginning from infinity at rest. In Newtonian mechanics, the energy of a particle is the sum of its kinetic and potential energy, and the sum is zero if the particle is at rest at infinity
\be
E = \frac{1}{2} m v^2 - G_{\rm N} \frac{M m}{r} \approx 0 \, ,
\ee
where $m$ and $v$ are the mass and the velocity of the particle falling onto the black hole and $M$ is the black hole mass. At the radial coordinate $r \sim 10$~$r_{\rm g}$, the kinetic energy of the particle is around 10\% of its rest mass, namely around 100~MeV for protons and 50~keV for electrons. We can thus expect the emission of radiation with such an energy, which is indeed in the X-ray and $\gamma$-ray bands.

A large part of the electromagnetic spectra is blocked by the Earth's atmosphere, see Fig.~\ref{f-atmosphere}. If it were not so, life on Earth -- at least as we know -- would be impossible, because $\gamma$-rays, X-rays, and UV photons are harmful for any organism. To be able to observe X-rays and $\gamma$-rays, observatories must thus be on board rockets or satellites. The first X-ray observatory can be considered a V2~rocket launched in 1948, which was used to observe the Sun, the brightest X-ray source in the sky. The first extrasolar X-ray source was discovered in 1962 by a team led by Riccardo Giacconi with an X-ray detector on board of an Aerobee 150 sounding rocket~\cite{r-intro-giacconi}. The source, known as Scorpius~X-1, is an X-ray binary with a neutron star of 1.4~$M_\odot$ and a companion of 0.42~$M_\odot$. Giacconi received the Nobel Prize in Physics in 2002 for pioneering the research field today called X-ray astronomy. Since the discovery of Scorpius~X-1, a steady progress in technology, theory and analysis, has made X-ray astronomy a leading scientific field in astrophysics research. Tabs.~\ref{tab-summary-xraymissions} and \ref{tab-summary-gammaraymissions} present some of the most important X-ray and $\gamma$-ray observatories from past, present, and future. 

\begin{table}[h]
\centering
{\renewcommand{\arraystretch}{1.15}
\begin{tabular}{||ccccccc||}
\hline
Mission & \hspace{0.1cm} & Launch Date & \hspace{0.1cm} & End of Mission & \hspace{0.1cm} & Instruments \\ 
\hline \hline
\multicolumn{7}{||l||}{\hspace{0.1cm}\textbf{PAST}}\\
\hline
R\"ontgensatellit (ROSAT) && 1990 && 1999 && XRT (0.1-2~keV) \\
\hline
Advanced Satellite for Cosmology && 1993 && 2000 && GIS (0.7-10~keV) \\
and Astrophysics (ASCA) && && && SIS (0.4-10~keV) \\
\hline
Rossi X-ray Timing Explorer && 1995 && 2012 && ASM (2-10~keV) \\
(RXTE) &&  &&  && PCA (2-60~keV)\\
&&  &&  && HEXTE (15-250~keV)\\
\hline
Suzaku && 2005 && 2015 && XRS (0.3-12~keV) \\
&&  &&  && XIS (0.2-12~keV) \\ 
&&  &&  && HXD (10-600~keV) \\
\hline
Hitomi && 2016 && 2016 && SXS (0.4-12~keV) \\
&& && && SXI (0.3-12~keV) \\
&& && && HXI (5-80~keV) \\
\hline
\multicolumn{7}{||l||}{\hspace{0.1cm}\textbf{PRESENT}}\\
\hline
Chandra X-ray Observatory && 1999 && -- && ACIS (0.2-10~keV)\\
(CXO) && && && HRC (0.1-10~keV)\\
&& && && LETG (0.08-2~keV) \\
&& && && HETG (0.4-10~keV) \\
\hline
XMM-Newton && 1999 && -- && EPIC-MOS (0.15-15~keV) \\ 
&& && && EPIC-pn (0.15-15~keV) \\ 
&& && && RGS (0.33-2.5~keV) \\
\hline
International Gamma-Ray && 2002 && -- && IBIS (15~keV-10~MeV) \\
Astrophysics Laboratory && && && SPI (18~keV-8~MeV) \\
(INTEGRAL) && && && JEM-X (3-35~keV) \\
\hline
Swift && 2004 && -- && BAT (15-150~keV) \\
 && && && XRT (0.2-10~keV) \\
\hline
Monitor of All-sky X-ray Image && 2009 && -- && SSC (0.5-10~keV) \\
(MAXI) && && && GSC (2-30~keV) \\
\hline
Nuclear Spectroscopic Telescope && 2012 && -- && FPMA (3-79~keV) \\
Array (NuSTAR) && && && FPMB (3-79~keV) \\
\hline
ASTROSAT && 2015 && -- && SXT (0.3-80~keV) \\
&& && && LAXPC (3-80~keV) \\
&& && && CZTI (100-300~keV) \\
\hline 
Neutron star Interior && 2017 && -- && XTI (0.2-12keV) \\
Composition Explorer (NICER) && && && \\
\hline
Hard X-ray Modulation Telescope && 2017 && -- && HE (20-250~keV) \\
(HXMT) && && && ME (5-30~keV) \\
&& && && LE (1-15~keV) \\
\hline
Spektrum-Roentgen-Gamma && 2019 && -- && eROSITA (0.3-10~keV) \\
(Spektr-RG) && && && ART-XC (0.5-11~keV) \\
\hline
\multicolumn{7}{||l||}{\hspace{0.1cm}\textbf{FUTURE}}\\
\hline
X-Ray Imaging and Spectroscopy && 2022 && -- && Resolve (0.4-12~keV) \\
Mission (XRISM) && && && Xtend (0.3-12~keV) \\
\hline
Enhanced X-ray Timing && 2027 && -- && SFA (0.5-20~keV) \\
Polarization (eXTP)  && && && LAD (1-30~keV) \\
\hline
Advanced Telescope for High && 2031 && -- && X-IFU (0.2-12~keV) \\
Energy Astrophysics (ATHENA) && && && WFI (0.1-15~keV) \\
\hline
\end{tabular}}
\vspace{0.4cm}
\caption{List of some of the most important X-ray missions from past, present, and future. \label{tab-summary-xraymissions}}
\end{table}

\begin{table}[h]
\centering
{\renewcommand{\arraystretch}{1.15}
\begin{tabular}{||ccccccc||}
\hline
Mission & \hspace{0.1cm} & Launch Date & \hspace{0.1cm} & End of Mission & \hspace{0.1cm} & Instruments \\ 
\hline \hline
\multicolumn{7}{||l||}{\hspace{0.1cm}\textbf{PAST}}\\
\hline
GRANAT && 1989 && 1999 && SIMGA (30-1300~keV) \\
&& && && PHEBUS (0.1-100~MeV) \\
&& && && KONUS-B (0.01-8~MeV) \\
&& && && TOURNESOL (0.002-20~MeV)\\
\hline
Compton Gamma Ray && 1991 && 2000 && OSSE (0.06-10~MeV) \\
Observatory (CGRO) &&  &&  && COMPTEL (0.8-30~MeV))\\
&&  &&  && EGRET (20-3000~MeV)\\
&& && && BATSE (0.015-110~MeV)\\
\hline
\multicolumn{7}{||l||}{\hspace{0.1cm}\textbf{PRESENT}}\\
\hline
International Gamma-Ray && 2002 && -- && SPI (0.02-8~MeV)\\
Astrophysics Laboratory && && && IBIS (0.015-10~MeV)\\
(INTEGRAL) && && &&\\
\hline
Swift && 2004 && -- && BAT (15-150~keV) \\
\hline
Astrorivelatore Gamma && 2007 && -- && GRID (30~MeV-50~GeV) \\
ad Immagini LEggero && && && MC (0.25-200~MeV)\\
(AGILE)  && && && \\ 
\hline
Fermi gamma ray && 2008 && -- && LAT (20~Mev-300~GeV) \\
space telescope && && && GBM (8~keV-30~MeV)\\
\hline
\end{tabular}}
\vspace{0.4cm}
\caption{List of some of the most important $\gamma$-ray missions from past and present. \label{tab-summary-gammaraymissions}}
\end{table}


\section{Open problems and future directions}\label{s1-prob}

X-ray and $\gamma$-ray radiation have provided invaluable information about black holes and their astrophysical environments and breakthroughs in fundamental physics. In the case of accreting black holes, we can study of the accretion process in the strong gravity region, how the gas falls onto the compact object, and how jets and outflows are generated. In the past 10-15~years, a few X-ray techniques have been developed to measure black hole spins, and before the detection of gravitational waves these were the only techniques capable of measuring black hole spins.

While several puzzles have been answered, many new ones have appeared and remain unresolved. Some of them are as follows:  
\begin{enumerate}
\item While Einstein's general relativity is the standard framework for describing the gravitational features in our Universe, several shortcomings of the theory have led to the development of a large number of modified theories of gravity. The techniques used for measuring black hole spin can also be used to test the motion of particle in the strong gravity region around black holes from modified theories of gravity and be used to perform precision tests of general relativity~\cite{r-intro-book,r-intro-revmodphys,r-intro-Cao:2017kdq,r-intro-Tripathi:2018lhx}. 
\item There are a number of dark matter models predicting the production of $\gamma$-rays from dark matter particle annihilation or decay. The study of the $\gamma$-ray spectrum of astrophysical sites where there may be a large amount of dark matter particles is an indirect search for dark matter. If we detect an excess of $\gamma$-rays with respect to that expected from the pure astrophysical environment, as well as some specific feature in the $\gamma$-ray spectrum, this may be interpreted as an indirect evidence of dark matter particles.
\item What is the spin distribution among stellar-mass and supermassive black holes and how does the spin distribution change over cosmological times? In the case of supermassive black holes, the spin distribution would also provide information about the evolution of their host galaxies~\cite{r-intro-berti08}.
\item What is the mechanism responsible for the production of jets in black holes?
\item What is the mechanism responsible of the observed quasi-periodic oscillations (QPOs) in the X-ray power density spectrum of black holes? Can we use QPOs for measuring black hole spins and test general relativity?
\item What is the exact origin of supermassive black holes and how do they grow so fast? In particular, we know supermassive black holes of billions of Solar masses at redshifts higher than 6 and we do not have a clear understanding of how such objects were created and were able to grow so fast in a relatively short time.
\item How does the host environment determine the properties of supermassive black holes? And how do supermassive black holes determine the properties of their host environment?
\item Do intermediate mass black holes exist? Do small primordial black holes created in the early Universe exist? 
\end{enumerate}

\end{document}